\documentclass{article}

\usepackage{arxiv}

\usepackage[utf8]{inputenc} 
\usepackage[T1]{fontenc}    
\usepackage{hyperref}       
\usepackage{url}            
\usepackage{booktabs}       
\usepackage{amsfonts}       
\usepackage{nicefrac}       
\usepackage{microtype}      
\usepackage{lipsum}
\usepackage{amsmath}
\usepackage{tikz, pgfplots} 

\title{Analytical solution of the Euler-Poinsot problem}

\author{
  Cássio Murakami\\
  Department of Mechanical Engineering\\
  Polytechnic School of USP\\
  São Paulo, SP\\
  \texttt{cassiomura@usp.br} 
}

\pgfplotsset{compat = 1.17}
\begin{document}
\maketitle

\begin{abstract}

In the present paper, an analysis was performed on the torque-free motion of a rigid body, developing Euler's analytical solution and Poinsot's geometric solution. From mathematical formulations, the analytical solution for the time evolution of the angular velocity and Euler's angles was obtained and described given some initial conditions. Besides, an animation of Poinsot's geometric solution was elaborated and a study was carried out on the conditions in which the herpolhode forms a closed curve. Finally, an algorithm was developed in the software \textit{Scilab} that displays the analytical and numerical solutions obtained, it also generates an animation of the geometric solution, moreover to having an algorithm that generates closed herpolhodes.

\end{abstract}


\section{Introduction}
The elaboration of models for the representation of physical situations is a valuable technique for the development of projects in several areas of knowledge, such as Engineering, Natural Sciences, Human Sciences, among others. The advancement of technology has provided mechanisms for the construction of models and due to the implementation of computational algorithms for the solution of differential systems, a wide range of problems that were previously restricted to the need for an analytical solution can then be simulated with precision.

Whereas modeling methods have become more intuitive, they can still have inconsistencies and it is often the user's responsibility to detect them. Such a task requires a rigorous and detailed analysis of the results obtained, to ensure that both the model was applied correctly, and that the computational solution method resulted in a coherent response.

However, as explained in \cite{hazelrigg2003thoughts}, modeling errors are much more frequent in scientific and engineering practice than might be supposed at first sight. Considering that these errors are often present in models implemented in computational tools to aid engineering, which, in turn, are used by a wide community, it is clear how serious the consequences of their non-detection can be. Besides, the computational solution is based on numerical techniques, which presents a restriction in the analysis of the problem, since a function that describes the evolution of the system is not obtained, but only the result for a given scenario. Therefore, the development of an analytic solution is useful both to check the consistency of the computational model \cite{sargent2010verification} and to offer a clear insight into the situation.

Moreover, the development of a reliable model can be useful in the teaching process even for classical mechanical problems \cite{andaloro1991modelling}. For instance, the motion of a torque-free rigid body, which even though it has several pieces of literature that approach it in different ways \cite{garnier1954cinematique}, \cite{meirovitch2010methods}, \cite{Goldstein}, some deepening can be carried out to further clarify the problem. To mention a few: elaborate an accessible algorithm that would make available the results obtained for the imposed scenario and animate or illustrate the physical situation \cite{zabunov2013effect}.

\section{Theoretical Background}

\subsection{The Euler-Poinsot Problem}
The Euler-Poinsot problem consists of the study of the motion of a rigid body that is not necessarily axisymmetric, which rotates freely (not subject to any net forces or torques) around a fixed point.

Assume a rigid body with a generic geometry, fixed at the point $O$ of space, and adopt a coordinate system $O_{xyz}$ attached to the rigid body, whose axes coincide with the principal axes of rotation relative to the point $O$ of the rigid body. Also, assume an inertial coordinate system $O_{XYZ}$ fixed in space.

Adopt the principal moments of inertia corresponding to the $x$, $y$, and $z$ axes of the coordinate system attached to the body being, respectively, $ I_{x}$, $I_{y}$, and $I_{z}$. The study will be carried out considering a non-symmetric rigid body, and without loss of generality, that $I_{x} > I_{y} > I_{z}$.

Let the angular velocity vector related to the moving base be described as
\begin{equation*}
    \Vec{\omega} = (\omega_{x}, \omega_{y}, \omega_{z}).
\end{equation*}
The angular momentum $\overrightarrow{H}_{o}$ of a rigid body in relation to the point $O$ is given by
\begin{equation}
    \overrightarrow{H}_{o} = M(G - O)  \times  \dot{\Vec{r}}_{o}
    + J_{Oxyz} \;\vec{\omega}
    \label{eq:momento_quantidade}
\end{equation}
where $M$ is the mass of the rigid body, $G$ the position of the center of mass, $\Vec{r}_{o}$ the position of the point $o$ chosen relative to the point $O$, and $J_{Oxyz}$ the inertia tensor of the rigid body.

From the angular momentum theorem
\begin{equation}
    \dfrac{\mathrm{d}\overrightarrow{H}_{o}}{\mathrm{d}t} = \overrightarrow{M}_{o}^{ext} - \dot{\Vec{r}}_{o}  \times  M\overrightarrow{V}_{G}
    \label{TQMA}
\end{equation}
where $\Vec{M_{o}^{ext}}$ is the net torque and $\overrightarrow{V}_{G}$ the velocity of the body's center of mass.

Differentiating the expression (\ref{eq:momento_quantidade}) and applying the result to the equation (\ref{TQMA}), the following equation is obtained

\begin{equation}
    \overrightarrow{M_{o}}^{ext} = M(G - O)\times  \ddot{\Vec{r}}_{o} + J_{Oxyz}\;\dot{\Vec{\omega}} + \Vec{\omega}\times(J_{Oxyz}\;\Vec{\omega}).
    \label{eq:equation_TQMA}
\end{equation}

The conditions of the physical situation to be studied are described below

\begin{itemize}
    \item[\textasteriskcentered] Point $O$ as a fixed point: $\ddot{\Vec{r}}_{o} = \Vec{0}$.
    \item[\textasteriskcentered] System without the action of net torques: $\overrightarrow{M}_{o}^{ext}$ = $\Vec{0}$.
    \item[\textasteriskcentered] Coordinate system $O_{xyz}$ coinciding with the principal axes of rotation: $J_{Oxyz} = \begin{bmatrix} I_{x}&0&0\\0&I_{y}&0\\0&0&I_{z}\end{bmatrix} \cdot$
\end{itemize}

Applying such conditions to the equation (\ref{eq:equation_TQMA}), the following system of differential equations is obtained

\begin{equation} \label{Euler}
\begin{split} 
        I_{x} \dot{\omega}_{x} &= (I_{y} - I_{z})\omega_{y} \omega_{z} \\
        I_{y} \dot{\omega}_{y} &= (I_{z} - I_{x})\omega_{x} \omega_{z} \\
        I_{z} \dot{\omega}_{z} &= (I_{x} - I_{y})\omega_{x} \omega_{y}. 
\end{split}
\end{equation}

Once the system of differential equations (\ref{Euler}) is solved, the time evolution of the angular velocity of the rigid body in relation to the coordinate axes fixed to the body will be obtained.

From the equations of the described differential system it is possible to obtain important expressions for the development of the problem:

\begin{itemize}
    \item \textbf{Kinect energy ($T$)}

    Multiplying the equations described in the system (\ref{Euler}), respectively, by $\omega_{x}$, $\omega_{y}$, $\omega_{z}$ and adding the three results

        \begin{equation*}
            I_{x}\omega_{x}\dot{\omega}_{x} + I_{y}\omega_{y}\dot{\omega}_{y} + I_{z}\omega_{z}\dot{\omega}_{z} = 0 \rightarrow I_{x}\int \omega_{x} \mathrm{d} \omega_{x} + I_{y}\int \omega_{y} \mathrm{d} \omega_{y} + I_{z}\int \omega_{z} \mathrm{d} \omega_{z} = 0.
        \end{equation*}

    Integrating the expression, the following relation is obtained

        \begin{equation}
            I_{x}\omega_{x}^{2} + I_{y}\omega_{y}^{2} + I_{z}\omega_{z}^{2} = 2T = \text{const}.
        \label{eq:cinetica}
        \end{equation}

    \item \textbf{Angular momentum ($G$)}

     Multiplying the equations described in the system (\ref{Euler}), respectively, by $I_{x}\omega_{x}$, $I_{y}\omega_{y}$, $I_{z}\omega_{z}$ and adding the three results
        \begin{equation*}
            I_{x}^{2}\omega_{x}\dot{\omega}_{x} + I_{y}^{2}\omega_{y}\dot{\omega}_{y} + I_{z}^{2}\omega_{z}\dot{\omega}_{z} = 0 \rightarrow I_{x}^{2}\int \omega_{x} \mathrm{d} \omega_{x} + I_{y}^{2}\int \omega_{y} \mathrm{d} \omega_{y} + I_{z}^{2}\int \omega_{z} \mathrm{d} \omega_{z} = 0.
        \end{equation*}

        Integrating the expression, the following relation is obtained

        \begin{equation}
            I_{x}^{2}\omega_{x}^{2} + I_{y}^{2}\omega_{y}^{2} + I_{z}^{2}\omega_{z}^{2} = G^{2} = \text{const}.
        \label{eq:qtnd_movi}
        \end{equation}

\end{itemize}
\subsection{Euler's Angles}

\begin{figure}[ht]
    \centering
    \includegraphics[width=50mm]{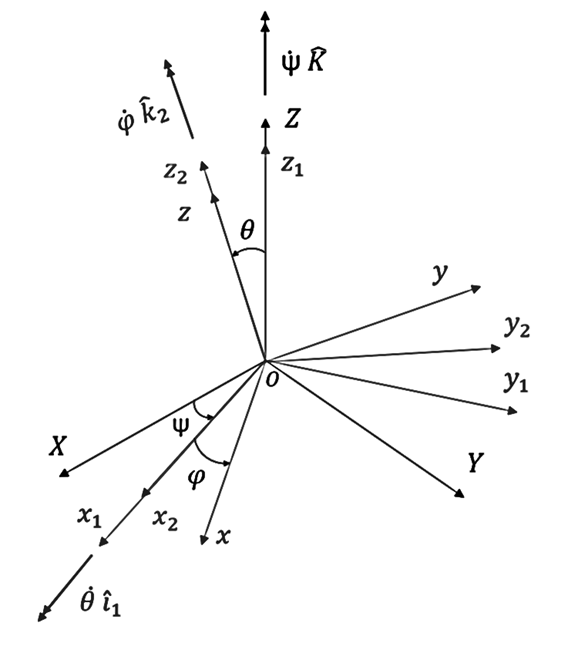}
    \caption{Euler's angles considering the rotation $Z-x_{1}-z_{2}$.}
    \label{Euler_angle}
\end{figure}

To study kinematics, the classical set of Euler's angles \cite{Mladenov} will be chosen, which consists of the rotation $ Z-x_{1}-z_{2}$. Fig. \ref{Euler_angle} illustrates the Euler's angles selected.

The rotation matrices corresponding to each step of the change of basis can be obtained from the direction cosine matrix transformation. To describe the rotation matrices, the following notation will be adopted: $\text{s}x$ for $\sin{x}$ and $\text{c}x$ for $\cos{x}$.

\begin{itemize}
    \item Precession $(\psi)$: $O_{XYZ} \rightarrow O_{x_{1}y_{1}z_{1}}, \qquad(Z = z_{1})$

    \begin{equation}
        \begin{bmatrix}
            \hat{i}_{1}\\
            \hat{j}_{1}\\
            \hat{k}_{1}
        \end{bmatrix} =
        \begin{bmatrix}
            \hat{I} \cdot \hat{i}_{1} & \hat{J} \cdot \hat{i}_{1}&\hat{K} \cdot \hat{i}_{1}\\
            \hat{I} \cdot \hat{j}_{1} & \hat{J} \cdot \hat{j}_{1} & \hat{K} \cdot \hat{j}_{1}\\
            \hat{I} \cdot \hat{k}_{1} & \hat{J} \cdot \hat{k}_{1} & \hat{K} \cdot \hat{k}_{1}
        \end{bmatrix}
        \begin{bmatrix}
            \hat{I}\\
            \hat{J}\\
            \hat{K}
        \end{bmatrix} =
        \begin{bmatrix}
            \text{c} \psi&\text{s}\psi&0\\
            -\text{s}\psi&\text{c}\psi&0\\
            0&0&1
        \end{bmatrix}
        \begin{bmatrix}
            \hat{I}\\
            \hat{J}\\
            \hat{K}
        \end{bmatrix}
        \label{transfor1}
    \end{equation}

    \item Nutation ($\theta$): $O_{x_{1}y_{1}z_{1}}\rightarrow O_{x_{2}y_{2}z_{2}},\qquad(x_{1} = x_{2})$

    \begin{equation}
        \begin{bmatrix}
            \hat{i}_{2}\\
            \hat{j}_{2}\\
            \hat{k}_{2}
        \end{bmatrix} =
        \begin{bmatrix}
            \hat{i}_{1} \cdot \hat{i}_{2} & \hat{j}_{1} \cdot \hat{i}_{2}&\hat{k}_{1} \cdot \hat{i}_{2}\\
            \hat{i}_{1} \cdot \hat{j}_{2} & \hat{j}_{1} \cdot \hat{j}_{2} & \hat{k}_{1} \cdot \hat{j}_{2}\\
            \hat{i}_{1} \cdot \hat{k}_{2} & \hat{j}_{1} \cdot \hat{k}_{2} & \hat{k}_{1} \cdot \hat{k}_{2}
        \end{bmatrix}
        \begin{bmatrix}
            \hat{i}_{1}\\
            \hat{j}_{1}\\
            \hat{k}_{1}
        \end{bmatrix} =
        \begin{bmatrix}
            1&0&0\\
            0&\text{c}\theta&\text{s}\theta\\
            0&-\text{s}\theta&\text{c}\theta
        \end{bmatrix}
        \begin{bmatrix}
            \hat{i}_{1}\\
            \hat{j}_{1}\\
            \hat{k}_{1}
        \end{bmatrix}
        \label{tranfor2}
    \end{equation}

    \item Intrinsic rotation ($\varphi$): $O_{x_{2}y_{2}z_{2}}\rightarrow O_{xyz},\qquad(z_{2} = z)$

        \begin{equation}
            \begin{bmatrix}
                \hat{i}\\
                \hat{j}\\
                \hat{k}
            \end{bmatrix} =
            \begin{bmatrix}
                \hat{i}_{2} \cdot \hat{i} & \hat{j}_{2} \cdot \hat{i}&\hat{k}_{2} \cdot \hat{i}\\
                \hat{i}_{2} \cdot \hat{j} & \hat{j}_{2} \cdot \hat{j} & \hat{k}_{2} \cdot \hat{j}\\
                \hat{i}_{2} \cdot \hat{k} & \hat{j}_{2} \cdot \hat{k} & \hat{k}_{2} \cdot \hat{k}
            \end{bmatrix}
        \begin{bmatrix}
            \hat{i}_{2}\\
            \hat{j}_{2}\\
            \hat{k}_{2}
        \end{bmatrix} =
        \begin{bmatrix}
            \text{c}\varphi&\text{s}\varphi&0\\
            -\text{s}\varphi&\text{c}{\varphi}&0\\
            0&0&1
        \end{bmatrix}
        \begin{bmatrix}
            \hat{i}_{2}\\
            \hat{j}_{2}\\
            \hat{k}_{2}
        \end{bmatrix} \cdot
    \label{tranfor3}
    \end{equation}
\end{itemize}

Once the transformation matrices for each step of the change of coordinates are known, it is possible to write the transformation matrix that takes from the fixed coordinate in space to the fixed coordinate in the moving body performing successive multiplications of the transformation matrices (\ref{tranfor3}), (\ref{tranfor2}), and (\ref{transfor1}). The result of this process is shown below

\begin{equation}
    \begin{bmatrix}
        \hat{i}\\
        \hat{j}\\
        \hat{k}
    \end{bmatrix} =
    \begin{bmatrix}
        \text{c}\psi \text{c}\varphi - \text{s}\psi \text{c}\theta\text{s}\varphi & \text{s}\psi \text{c}\varphi + \text{c}\psi\text{c}\theta\text{s}\varphi & \text{s}\theta\text{s}\varphi\\
        -\text{c}\psi \text{s}\varphi - \text{s}\psi \text{c}\theta \text{c} \varphi & -\text{s}\psi \text{s}\varphi + \text{c}\psi \text{c}\theta \text{c} \varphi & \text{s}\theta\text{c}\varphi\\
         \text{s} \psi \text{s}\theta & -\text{c}\psi \text{s}\theta & \text{c}\theta
    \end{bmatrix}
    \begin{bmatrix}
        \hat{I}\\
        \hat{J}\\
        \hat{K}
    \end{bmatrix} \cdot
    \label{trans1}
\end{equation}

\section{Analytical Solutions}

\subsection{Analytical solution of the angular velocity}

The analytical solution of the angular velocity consists in solving the differential system that defines the Euler-Poinsot problem given by

\begin{equation} \label{Euler_cases}
\begin{split}
        I_{x} \dot{\omega}_{x} &= (I_{y} - I_{z})\omega_{y} \omega_{z} \\
        I_{y} \dot{\omega}_{y} &= (I_{z} - I_{x})\omega_{x} \omega_{z} \\
        I_{z} \dot{\omega}_{z} &= (I_{x} - I_{y})\omega_{x} \omega_{y}\\
\end{split},
\qquad
\begin{split}
    \omega_{x}(0) &= \omega_{x0}\\
    \omega_{y}(0) &= \omega_{y0}\\
     \omega_{z}(0) &= \omega_{z0}.
\end{split}
\end{equation}

It is worth noting that from (\ref{eq:cinetica}) and (\ref{eq:qtnd_movi}) the constants $T$ and $G$ are determined by the initial conditions imposed

\begin{equation*}
    2T = I_{x}\omega_{x0}^{2} + I_{y}\omega_{y0}^{2} + I_{z}\omega_{z0}^{2},\qquad G^{2} = I_{x}^{2}\omega_{x0}^{2} + I_{y}^{2}\omega_{y0}^{2} + I_{z}^{2}\omega_{z0}^{2}.
    \label{cond_in}
\end{equation*}

The solution of the general case of the Euler-Poinsot problem has its analytical form imposed by elliptical functions \cite{greenhill1892applications}. To obtain the analytical solution, the problem will be divided into two cases.

$\triangleright \quad$ \textbf{First case: $2TI_{x} > 2TI_{y} > G^{2} > 2TI_{z}$}

    The solution takes the form

    \begin{equation}
        \omega_{x}(t) = P \;\text{cn}(nt + \tau,k), \qquad \omega_{y}(t) = -Q \;\text{sn}(nt + \tau,k), \qquad \omega_{z}(t) =  R \;\text{dn}(nt + \tau,k)
        \label{sol_forma}
    \end{equation}
  
    where $P,Q,R,n,k\;\text{and}\;\tau$ are constants that will be determined.

    Differentiating and replacing the expressions (\ref{sol_forma}) in the system of differential equations (\ref{Euler_cases}), and upon rearrangement, the following expressions are obtained
    
    \begin{equation} \label{eq: relação_1}
        \dfrac{I_{y}-I_{z}}{I_{x}} = \dfrac{nP}{QR},\qquad
        \dfrac{I_{x}-I_{z}}{I_{y}} = \dfrac{nQ}{PR},\qquad
        \dfrac{I_{x}-I_{y}}{I_{z}} = \dfrac{k^{2}nR}{PQ} \cdot
    \end{equation}

    In addition to these relations, the solution must be valid for any instant of time, including the instant $t' = -\frac{\tau}{n}$. Analyzing that instant in the functions (\ref{sol_forma})

    \begin{equation*}
    \omega_{x}(t') = P,\qquad \omega_{y}(t') = 0,\qquad \omega_{z}(t') = R.
    \end{equation*}

    Therefore, considering the instant $t'$, the kinetic energy (\ref{eq:cinetica}) and the square of the absolute value of the angular momentum vector (\ref{eq:qtnd_movi}) are given by
    \begin{equation*}
        \begin{split}
            I_{x}P^{2} + I_{z}R^{2} = 2T,\qquad
            I_{x}^{2}P^{2} + I_{z}^{2}R^{2} = G^{2}. 
        \end{split}
    \end{equation*}
    Multiplying the kinect energy equation by $I_{z}$ and subtracting the result from the angular momentum relation, the following value for $P^2$ is obtained
    \begin{equation*}
        P^{2} = \frac{G^{2} - 2TI_{z}}{I_{x}^{2} - I_{x}I_{z}} \cdot
    \label{P_1}
    \end{equation*}

    Multiplying the kinect energy equation by $I_{x}$ and subtracting the result from the angular momentum relation, the following value for $R^2$ is obtained
    \begin{equation*}
        R^{2} = \frac{G^{2} - 2TI_{x}}{I_{z}^{2} - I_{x}I_{z}} \cdot
    \label{R_1}
    \end{equation*}

    Upon algebraic manipulation of equations (\ref{eq: relação_1}) and replacing the value of $P^2$, the following value for $Q^2$ is obtained
    \begin{equation*}
        Q^{2} = \frac{G^{2} - 2TI_{z}}{I_{y}^{2} - I_{y}I_{z}} \cdot
        \label{Q_1}
    \end{equation*}

    Upon algebraic manipulation of equations (\ref{eq: relação_1}) and replacing the value of $R^2$, and considering $n\geq 0$ the following value for $n$ is obtained
    \begin{equation}
        n = \sqrt{\frac{(I_{y}-I_{z})(2TI_{x} - G^{2})}{I_{x}I_{y}I_{z}}} \cdot
    \label{n_caso1}
    \end{equation}

    Upon algebraic manipulation of equations (\ref{eq: relação_1}) and replacing the values of $P^2$ and $R^2$ obtained, and considering $k\geq0$ the following value is obtained for $k$
    \begin{equation}
            k = \sqrt{\frac{I_{x}-I_{y}}{I_{y}-I_{z}}\;\frac{G^{2}- 2TI_{z}}{2TI_{x}-G^{2}}} \cdot
    \label{k_caso1}
    \end{equation}

    As the expressions obtained for the constants are quadratic, it is necessary to carry out an analysis of their signs in (\ref{sol_forma}). That choice is directly linked to the $\omega_{z0}$ sign. For instance, the function $\text{dn}(nt + \tau, k)$ is strictly positive and if the sign of $\omega_{z0}$ is chosen to be negative, the value of $R$ must be necessarily negative. Also, considering $Q > 0$, which is justified by the constant $\tau$ that will be obtained for a positive $Q$, from equation (\ref{eq: relação_1}) the sign of $P$ must be the same as the sign of $R$, to the relations be valid. Such a study can be separated into two cases:

    \begin{itemize}
        \item If $\omega_{z0} \geq 0$

        \begin{equation*}
            P = \sqrt{\frac{G^{2} - 2TI_{z}}{I_{x}^{2} - I_{x}I_{z}}}, \qquad Q = \sqrt{\frac{G^{2} - 2TI_{z}}{I_{y}^{2} - I_{y}I_{z}}}, \qquad R = \sqrt{\frac{G^{2} - 2TI_{x}}{I_{z}^{2} - I_{x}I_{z}}}
        \end{equation*}

        \item If $\omega_{z0} < 0$

        \begin{equation*}
        P = -\sqrt{\frac{G^{2} - 2TI_{z}}{I_{x}^{2} - I_{x}I_{z}}}, \qquad Q = \sqrt{\frac{G^{2} - 2TI_{z}}{I_{y}^{2} - I_{y}I_{z}}}, \qquad R = -\sqrt{\frac{G^{2} - 2TI_{x}}{I_{z}^{2} - I_{x}I_{z}}} \cdot
        \end{equation*}

    \end{itemize}

    It is worth noting that these cases can be replaced by a unique formula by applying the signum function sgn($x$). Which is defined as follows: if $x > 0$ then sgn($x$) $= 1$, if $x<0$ then sgn($x$) $ = -1$, if $x = 0$ then sgn($x$) $= 0$.

   Finally, the constant $\tau$ will be chosen such as the initial condition imposed are valid. It is worth noting that all the initial conditions relations are valid for the same $\tau$. For this purpose, the equation $\omega_{y}(0)=\omega_{y0}$ will be evaluated and the following relation must be valid
   
    \begin{equation*}
        -\sqrt{\frac{G^{2} - 2TI_{z}}{I_{y}^{2} - I_{y}I_{z}}}\text{sn}(\tau,k) = \omega_{y0}.
   \end{equation*}
 
    The solution will be divided into cases according to the initial conditions of the angular velocity components since the value of $\tau$ is obtained by inverting the elliptical function. It is possible to notice that in a period $K(k)$ there are two coincident values for the elliptic function. The choice of the correspondent to the studied situation depends whether the initial value is contained in the increasing or decreasing interval of the elliptical function. This choice is determined by the sign of the initial value of $\dot{\omega_{y}}$, which by (\ref{Euler_cases}) is imposed by the product $\omega_{x0}\;\omega_{z0}$.

     Thus, translating the function to obtain the expected value using the inverse of elliptic function, $\tau$ is given by

    \begin{itemize}

    \item If $\omega_{x0}\;\omega_{z0} \geq0$
        
        \begin{equation}    \label{tau_11}
            \tau = \int_{0}^{\tau_{0}} \dfrac{\mathrm{d}u}{\sqrt{(1-u^{2})(1-k^{2}u^{2})}}, \qquad \tau_{0} = -\frac{\omega_{y0}}{\sqrt{\frac{G^{2} - 2TI_{z}}{I_{y}^{2} - I_{y}I_{z}}}}
        \end{equation}
    
    \item If $\omega_{x0}\;\omega_{z0} < 0$
        
        \begin{equation}    \label{tau_12}
            \tau = \frac{K(k)}{2} - \int_{0}^{\tau_{0}} \dfrac{\mathrm{d}u}{\sqrt{(1-u^{2})(1-k^{2}u^{2})}}, \qquad \tau_{0} = -\frac{\omega_{y0}}{\sqrt{\frac{G^{2} - 2TI_{z}}{I_{y}^{2} - I_{y}I_{z}}}}
        \end{equation}
        
    where $K(k)$ is the period of the elliptic function $\text{sn}(t,k)$ given by

    \begin{equation}
        K(k) = 4\int_{0}^{\frac{\pi}{2}} \dfrac{\mathrm{d}u}{\sqrt{1 - k^2 \sin^{2}u}} \cdot
    \label{periodo}
    \end{equation}
    \end{itemize}

    Thus, once all the constants are obtained and considering (\ref{n_caso1}), (\ref{k_caso1}), (\ref{tau_11}) or (\ref{tau_12}), the analytical solution for the system (\ref{Euler_cases}) if $2TI_{y} > G^{2}$ is as follows
 
    \begin{equation}    \label{cases1}
        \begin{split}
        \omega_{x}(t) &= \text{sgn}(\omega_{z0})\sqrt{\frac{G^{2} - 2TI_{z}}{I_{x}^{2} - I_{x}I_{z}}}\text{cn}(nt + \tau,\;k)\\
        \omega_{y}(t) &= -\sqrt{\frac{G^{2} - 2TI_{z}}{I_{y}^{2} - I_{y}I_{z}}}\text{sn}(nt + \tau,k)\\
        \omega_{z}(t) &= \text{sgn}(\omega_{z0})\sqrt{\frac{G^{2} - 2TI_{x}}{I_{z}^{2} - I_{x}I_{z}}}\;\text{dn}(nt + \tau,\;k).
        \end{split}
    \end{equation}

$\triangleright \quad$ \textbf{Second case:} $2TI_{x}> G^{2} > 2TI_{y}>2TI_{z}$

    The solution takes the form
    
    \begin{equation*}
        \omega_{x}(t) = P\; \text{dn}(nt + \tau,k), \qquad \omega_{y}(t) = -Q\; \text{sn}(nt + \tau,k), \qquad \omega_{z}(t) =  R\; \text{cn}(nt + \tau,k)
        \label{sol_forma_2}
    \end{equation*}
    where $P,Q,R,n,k\;\text{and}\;\tau$ are constants that are  determined in an analogous way to the previous case:

\begin{equation}        \label{caso2}
    \begin{gathered}
        P^{2} = \frac{2TI_{z} - G^{2}}{I_{x}I_{z} - I_{x}^{2}}, \qquad Q^{2} = \frac{2TI_{x}-G^{2}}{I_{x}I_{y} - I_{y}^{2}}, \qquad R^{2} = \frac{2TI_{x}-G^{2}}{I_{x}I_{z} - I_{z}^{2}}\\
        n = \sqrt{\frac{(I_{x}-I_{y})(G^{2} - 2TI_{z})}{I_{x}I_{y}I_{z}}},\qquad k = \sqrt{\frac{(I_{y}-I_{z})}{(I_{x}-I_{y})}\; \frac{(G^{2} - 2TI_{x})}{(2TI_{z} -G^{2})}} \cdot
    \end{gathered}
\end{equation}

 The values of $ P,Q,$ and $R$ will also be directly influenced by the initial value of the component that accompanies the elliptic function $\text{dn}(nt + \tau,k)$, which in this case is $P$. As in the previous case it will be considered a $Q > 0$ and $R$ must have the same sign of $P$ to satisfy equations (\ref{eq: relação_1}). Then
 
    \begin{equation*}
    P = \text{sgn}(\omega_{x0})\sqrt{\frac{2TI_{z} - G^{2}}{I_{x}I_{z} - I_{x}^{2}}},\qquad Q = \sqrt{\frac{2TI_{x}-G^{2}}{I_{x}I_{y} - I_{y}^{2}}},\qquad R = \text{sgn}(\omega_{x0})\sqrt{\frac{2TI_{x}-G^{2}}{I_{x}I_{z} - I_{z}^{2}}}
    \end{equation*}

      Finally, the value of the constant $\tau$ must satisfies the initial conditions, which is described by the following cases

    \begin{itemize}

    \item If $\omega_{x0}\;\omega_{z0} \geq0$

        \begin{equation}    \label{tau_21}
            \tau = \int_{0}^{\tau_{0}} \dfrac{\mathrm{d}u}{\sqrt{(1-u^{2})(1-k^{2}u^{2})}}, \qquad \tau_{0} = -\frac{\omega_{y0}}{\sqrt{\frac{2TI_{x}-G^{2}}{I_{x}I_{y} - I_{y}^{2}}}}
        \end{equation}
    
    \item If $\omega_{x0}\;\omega_{z0} < 0$
        \begin{equation}    \label{tau_22}
            \tau = \frac{K(k)}{2} - \int_{0}^{\tau_{0}} \dfrac{\mathrm{d}u}{\sqrt{(1-u^{2})(1-k^{2}u^{2})}}, \qquad \tau_{0} = -\frac{\omega_{y0}}{\sqrt{\frac{2TI_{x}-G^{2}}{I_{x}I_{y} - I_{y}^{2}}}}
        \end{equation}
        
     where $K(k)$ is the period of the elliptic function $\text{sn}(t,k)$ described in (\ref{periodo}).

\end{itemize}
      Once all the constants are obtained considering (\ref{caso2}), (\ref{tau_21}) or (\ref{tau_22}), the analytical solution for the system (\ref{Euler_cases}) if $G^{2} > 2TI_{y}$ is as follows

    \begin{equation} \label{cases2}
        \begin{split}
        \omega_{x}(t) &= \text{sgn}(\omega_{x0})\sqrt{ \frac{2TI_{z} - G^{2}}{I_{x}I_{z} - I_{x}^{2}}}\text{dn}(nt + \tau,\;k)\\
        \omega_{y}(t) &= -\sqrt{\frac{2TI_{x}-G^{2}}{I_{x}I_{y} - I_{y}^{2}}}\text{sn}(nt + \tau,k)\\
        \omega_{z}(t) &= \text{sgn}(\omega_{x0})\sqrt{\frac{2TI_{x}-G^{2}}{I_{x}I_{z} - I_{z}^{2}}}\;\text{cn}(nt + \tau,\;k).
        \end{split}
    \end{equation}

\subsection{Analytical Solution of the Euler's Angles}

The determination of the analytical solution of the Euler's angles \cite{landau1976mechanics} is essential to stipulate the motion of the body in relation to an inertial frame of reference. Thus it is an essential result to obtain the motion of Poinsot's geometric solution \cite{Goldstein}. As the position of the coordinate axes fixed in space is arbitrary, a coordinate system will be chosen such that the $Z$ axis contains the invariant angular momentum vector $\overrightarrow{H}_{o}$, also it will be considered a null initial precession angle $\psi_{0}$.

Using the transformation matrix (\ref{trans1}), it is possible to establish a relationship between the coordinates of the angular momentum vector described in the coordinates of the system attached to the rigid body and the coordinates of the angular momentum vector described in the coordinates of the fixed base in space

\begin{equation}
    \begin{bmatrix}
        \text{c}\psi \text{c}\varphi - \text{s}\psi \text{c}\theta\text{s}\varphi & \text{s}\psi \text{c}\varphi + \text{c}\psi\text{c}\theta\text{s}\varphi & \text{s}\theta\text{s}\varphi\\
        -\text{c}\psi \text{s}\varphi - \text{s}\psi \text{c}\theta \text{c} \varphi & -\text{s}\psi \text{s}\varphi + \text{c}\psi \text{c}\theta \text{c} \varphi & \text{s}\theta\text{c}\varphi\\
        \text{s} \psi \text{s}\theta & -\text{c} \psi \text{s}\theta & \text{c}\theta
    \end{bmatrix}
    \begin{bmatrix}
        0\\0\\G
    \end{bmatrix}
    =
    \begin{bmatrix}
        H_{O x}\\
        H_{O y}\\
        H_{O z}
    \end{bmatrix} \cdot
    \label{sistema_angu}
\end{equation}

Upon rearrangement, the system (\ref{sistema_angu}) gives
\begin{equation}    \label{eq: sol_sistema_angu}
    \begin{split}
    G\sin\theta\sin\varphi &= I_{x}\omega_{x}\\
    G\sin\theta\cos\varphi &= I_{y}\omega_{y}\\
    G\cos\theta &= I_{z}\omega_{z}.
    \end{split}
\end{equation}

Combining the equations of the system (\ref{eq: sol_sistema_angu}), the following result is obtained for Euler's angles of nutation and intrinsic rotation

\begin{equation}
    \cos\theta(t) = \frac{I_{z}\omega_{z}(t)}{G}, \qquad \tan\varphi(t) = \frac{I_{x}\omega_{x}(t)}{I_{y}\omega_{y}(t)} \cdot
    \label{trig_euler}
\end{equation}

It is still necessary to verify the inversion of the cosine and tangent functions to obtain the Euler's angles corresponding to the physical situation. Consider a nutation angle restricted to the domain $[0,\;\pi]$, thus the inversion of $\cos \theta$ is given by the function $\arccos$. On the other hand, by the geometrical representation of Euler's angles, the intrinsic rotation can be interpreted as the angle of the projection of the angular momentum vector $\overrightarrow{H}_{O}$ in $xy$ plane. To validate not only the relation (\ref{trig_euler}) but the physical interpretation where the intrinsic rotation angle can vary from $-\pi$ to $\pi$, $\varphi$ will be considered as the angle equals to the phase of the complex number $\zeta(t) = I_{y}\omega_{y}(t) + \mathrm{i}I_{x}\omega_{x}(t)$ for a real $t$. Therefore, $\varphi(t) = \text{arg}(\zeta(t))$ and to obtain the argument of $\zeta(t)$ the function atan2 will be applied. Which is defined as follows: if $x>0$ then $\text{atan2}(y,x) = \text{arctan}(\frac{y}{x})$, if $x<0$ and $y\geq0$ then $\text{atan2}(y,x) = \text{arctan}(\frac{y}{x}) + \pi$, if $x<0$ and $y<0$ then $\text{atan2}(y,x) = \text{arctan}(\frac{y}{x}) - \pi$, if $x=0$ and $y>0$ then $\text{atan2}(y,x) =+ \frac{\pi}{2}$, if $x=0$ and $y<0$ then $\text{atan2}(y,x) = -\frac{\pi}{2}$, and if $x=0$ and $y=0$ then $\text{atan2}(y,x) \text{ is undefined}$.

    \begin{equation}
         \theta(t) = \arccos\left(\frac{I_{z}\omega_{z}(t)}{G}\right),\qquad
         \varphi(t) = \text{atan2}(I_{x}\omega_{x}(t),I_{y}\omega_{y}(t)).
    \label{euler_sol_1}
    \end{equation}

Finally, as the analytical solution of the components of the angular velocity vector in the base attached to the body has already been determined in the form of elliptical functions, the result for the Euler's angles (\ref{euler_sol_1}) is obtained by using the solutions acquired in (\ref{cases1}) or (\ref{cases2}).

It is worth noting that as the time evolution of the Euler's angles of nutation and intrinsic rotation is given using elliptic functions, such Euler's angles will be periodic at least in a period $K(k)$, corresponding to the period of these elliptical functions.

In order to obtain the analytic solution for the precession angle $\psi(t)$, it will be necessary to first relate the angular velocity to the Euler's angles. From Fig. \ref{Euler_angle} it is possible to establish such a connection between the variables mentioned using the following relationship

\begin{equation}
    \Vec{\omega} = \dot{\psi}\;\hat{K} + \dot{\theta} \;\hat{i}_{1} + \dot{\varphi}\;\hat{k}_{2}.
    \label{vet_cru}
\end{equation}

Using the transformation matrices (\ref{transfor1}), (\ref{tranfor2}) and (\ref{tranfor3}) it is possible to express the unit vectors of the equation (\ref{vet_cru}) using the unit vectors of the system of coordinate attached to the body

\begin{equation*}
    \hat{K} = \sin\theta \sin\varphi\;\hat{i} + \sin\theta \cos\varphi\;\hat{j} + \cos\theta\; \hat{k}, \qquad
    \hat{i}_{1} = \cos\varphi\;\hat{i} - \sin\varphi\;\hat{j},\qquad \hat{k}_{2} = \hat{k}.
\end{equation*}

Therefore, upon rearrangement, the angular velocity vector described in the system attached to the rigid body is described by

\begin{equation*}
    \Vec{\omega} = (\dot{\psi}\sin\theta\sin\varphi + \dot{\theta}\cos\varphi)\;\hat{i} + (\dot{\psi} \sin\theta\cos\varphi - \dot{\theta}\sin\varphi)\;\hat{j} + (\dot{\psi}\cos\theta + \dot{\varphi})\;\hat{k}.
    \label{omega_euler}
\end{equation*}

Thus to obtain the solution it is necessary to solve the following system for $\dot{\psi}$

\begin{equation} \label{eq: sistema_analitico}
    \begin{split}
        \omega_{x} &=\dot{\psi} \sin\theta\sin\varphi + \dot{\theta}\cos\varphi \\
        \omega_{y} &=\dot{\psi}\sin\theta\cos\varphi - \dot{\theta}\sin\varphi \\
        \omega_{z} &=\dot{\psi}\cos\theta + \dot{\varphi}.
    \end{split}
\end{equation}
Considering the system (\ref{eq: sistema_analitico}), multiplying $\omega_{x}$ by $\sin\varphi$, $\omega_{y}$ by $\cos\varphi$ and adding the results

\begin{equation}
    \dot{\psi}(t) = \frac{\omega_{x}(t)\sin\varphi(t) + \omega_{y}(t)\cos{\varphi(t)}}{\sin{\theta(t)}} \cdot
    \label{quadratura}
\end{equation}

As the functions in (\ref{trig_euler}) were obtained, it is possible to simplify the equation (\ref{quadratura}). To manipulate trigonometric functions an analysis of the inverse functions defined in (\ref{euler_sol_1}) will be performed. As the function $\arccos$ has its image restricted to [0,\;$\pi$], it is possible to conclude that $\sin\theta \geq 0$, thus applying the Pythagorean identity, $\sin\theta$ will carry the positive sign. Besides, the function atan2 is defined as a phase of a complex number, thus it contemplates all the possible angles value from $-\pi$ to $\pi$. Then

\begin{equation*}
    \sin \theta = \frac{\sqrt{I_{x}^2\omega_{x}^2 + I_{y}^2\omega_{y}^2}}{G},\qquad\sin \varphi =  \frac{I_{x}\omega_{x}}{\sqrt{I_{x}^2\omega_{x}^2 + I_{y}^2\omega_{y}^2}},\qquad \cos \varphi = \frac{I_{y}\omega_{y}}{\sqrt{I_{x}^2\omega_{x}^2 + I_{y}^2\omega_{y}^2}} \cdot
\end{equation*}

Using the trigonometric results obtained, it is possible to simplify the expression obtained in (\ref{quadratura}) as

\begin{equation}
    \dot{\psi}(t) = G \frac{I_{x}\omega_{x}^{2}(t) + I_{y}\omega_{y}^{2}(t)}{I_{x}^{2}\omega_{x}^{2}(t) + I_{y}^{2}\omega_{y}^{2}(t)} \cdot
    \label{real_quad}
\end{equation}

To obtain the analytical solution of $\psi(t)$, the analytical results obtained for the components of the angular velocity (\ref {cases1}) and (\ref{cases2}) will be used. Thus, it will be necessary to divide the problem into two cases.

    $\triangleright \quad$ \textbf{First case:} $2TI_{x}>2TI_{y} > G^{2}>2TI_{z}$

    From the previous solutions obtained for the component of angular velocity for this scenario, it is concluded that the squares of the solutions obtained in (\ref{cases1}), are given by the following expressions
    \begin{equation*}
        \omega_{x}^{2}(t) = \frac{2TI_{z} - G^{2}}{I_{x}(I_{z} - I_{x})}\text{cn}^{2}(nt + \tau,k),\qquad \omega_{y}^{2}(t) = \frac{2TI_{z} - G^{2}}{I_{y}(I_{z}-I_{y})}\text{sn}^{2}(nt + \tau,k).
    \end{equation*}

    Substituting those results in the function (\ref{real_quad}), using the algebraic property of the elliptic functions and factoring to simplify the expression, the following result is obtained

    \begin{equation*}
        \dot{\psi}(t) = G\frac{I_{z} - I_{y} - (I_{x} - I_{y})\text{sn}^{2}(nt + \tau,k)}{I_{x}I_{z} - I_{x}I_{y} - I_{z}(I_{x}-I_{y})\text{sn}^{2}(nt + \tau,k)}\cdot
        \label{analogiazinha}
    \end{equation*}

    Multiplying and dividing the expression by $I_{z}$, then adding and subtracting $I_{x}I_{z}$ and $ I_{x}I_{y}$ in the numerator and factoring out the expression

    \begin{equation*}
        \dot{\psi}(t) = G \frac{I_{x}(I_{z}-I_{y}) - I_{z}(I_{x} - I_{y})\text{sn}^{2}(nt + \tau,k) + (I_{z}-I_{y})(I_{z} - I_{x})}{I_{x}I_{z}(I_{z}-I_{y}) - I_{z}^{2}(I_{x} - I_{y})\text{sn}^{2}(nt + \tau,k)}\cdot
    \end{equation*}

    The expression obtained can be described by the following sum of fractions

    \begin{equation*}
        \dot{\psi}(t) =\frac{G}{I_{z}} + \frac{G(I_{z}-I_{y})(I_{z}-I_{x})}{I_{x}I_{z}(I_{z}-I_{y}) - I_{z}^{2}(I_{x} - I_{y})\text{sn}^{2}(nt + \tau,k)}\cdot
    \end{equation*}

    Upon rearrangement and integration of the expression, the following result is obtained

    \begin{equation*}
        \psi(t) =\frac{G}{I_{z}}t - \frac{G(I_{x}-I_{z})}{I_{x}I_{z}}\int_{0}^{t}\frac{\mathrm{d}t}{1 + \frac{I_{z}(I_{x} - I_{y})}{I_{x}(I_{y}-I_{z})}\text{sn}^{2}(nt + \tau,k)}\cdot
        \label{result_pre1}
    \end{equation*}

    Finally, from a manipulation of the integral by substitution, the obtained function can be written in the form of an incomplete elliptic integral of the third kind \cite{byrd2013handbook}.

    \begin{equation}
        \psi(t) =  \frac{G}{I_{z}}t - \frac{G(I_{x} - I_{z})}{I_{x}I_{z}n} \text{\Large $\Pi$}\left(\text{am}(nt + \tau) ,\;-\tfrac{I_{z}(I_{x} - I_{y})}{I_{x}(I_{y} - I_{z})},\;k\right) + \Lambda
    \label{precession_sol_1}
    \end{equation}

    where the constants $n$, $k$ and $\tau$ are defined respectively  in (\ref{n_caso1}), (\ref{k_caso1}), and (\ref{tau_11}) or (\ref{tau_12}). The constant $\Lambda$ is a defined as
 
    \begin{equation*}
        \Lambda = \frac{G(I_{x} - I_{z})}{I_{x}I_{z}n} \int_{0}^{\tau}\frac{\mathrm{d}u}{1 + \frac{I_{z}(I_{x} - I_{y})}{I_{x}(I_{y}-I_{z})}\text{sn}^2(u,k)} \cdot
    \end{equation*}

    $\triangleright \quad$ \textbf{Second case:} $2TI_{x} > G^{2} > 2TI_{y} > 2TI_{z}$

    From the solutions obtained for this scenario, it is concluded that the squares of the solutions with the due constants, already defined in (\ref{cases2}), are given by the following expressions
    \begin{equation*}
        \omega_{x}^{2}(t) = \frac{2TI_{z} - G^{2}}{I_{x}I_{z} - I_{x}^{2}}\text{dn}^{2}(nt + \tau,k),\qquad \omega_{y}^{2}(t) = \frac{2TI_{x}-G^{2}}{I_{x}I_{y} - I_{y}^{2}}\text{sn}^{2}(nt + \tau,k).
    \end{equation*}

    In an analogous way to the development of the previous case, the following result is obtained

     \begin{equation*}
        \psi(t) =\frac{G}{I_{z}}t - \frac{G(I_{x}-I_{z})}{I_{x}I_{z}}\int_{0}^{t}\frac{\mathrm{d}t}{1 + \frac{I_{z}(I_{x} - I_{y})}{I_{x}(I_{y}-I_{z})}k^{2}\text{sn}^{2}(nt + \tau,k)} \cdot
    \end{equation*}

    Finally, from a manipulation of the integral by substitution, the obtained function can be written in the form of an incomplete elliptic integral of the third kind.

    \begin{equation}
        \psi(t) =  \frac{G}{I_{z}}t - \frac{G(I_{x} - I_{z})}{I_{x}I_{z}n} \text{\Large $\Pi$} \left(\text{am}(nt + \tau) ,\;-\tfrac{I_{z}(I_{x} - I_{y})}{I_{x}(I_{y} - I_{z})}k^{2},\;k\right) + \Lambda
    \label{precession_sol_2}
    \end{equation}

    where the constants $n$, $k$ and $\tau$ are defined in (\ref{caso2}), and (\ref{tau_21}) or (\ref{tau_22}). The constant $\Lambda$ is a defined as
    
    \begin{equation*}
        \Lambda = \frac{G(I_{x} - I_{z})}{I_{x}I_{z}n} \int_{0}^{\tau}\frac{\mathrm{d}u}{1 + \frac{I_{z}(I_{x} - I_{y})}{I_{x}(I_{y}-I_{z})}k^{2}\text{sn}^2(u,k)} \cdot
    \end{equation*}

It is worth mentioning that the form of $\psi(t)$ is not an elliptic function of the first kind like those used to define the time evolution of the angular velocity vector. Thus, it is not expected that the time evolution of precession will have a period $K(k)$ as the other Euler's angles.

\subsubsection{Closed Herpolhode}

The difference between the period of functions $\theta(t)$ and $\varphi(t)$ in relation to $\psi(t)$ results in a time evolution of the herpolhode that is not necessarily repeated at each period of the functions mentioned. For the herpolhode to be traced so that it forms a closed curve, the motion of the Poinsot's ellipsoid needs to be repeated, to this purpose, it is necessary to have a synchronization between the periods of precession with those of nutation and intrinsic rotation, which becomes possible if after a certain interval of time the three angles simultaneously repeat their states.

One way to impose this phenomenon is to make the value of the precession angle after $K(k)$ defined by elliptic functions a multiple of $2\pi$ \cite{analytic_formula}. If such a situation occurs, when the $K(k)$ period is completed, the precession will be starting a new cycle, which synchronizes the movements.

The value of the precession angle after a period $K(k)$ is obtained by analyzing the analytical solutions obtained for the precession angle (\ref{precession_sol_1}) and (\ref{precession_sol_2}) at time $K(k)$.

$\triangleright \quad$ \textbf{First case: $2TI_{x} > 2TI_{y} > G^{2} > 2TI_{z}$}

        \begin{equation}
        \psi(K(k)) =  \frac{G}{I_{z}}K(k) - \frac{G(I_{x} - I_{z})}{I_{x}I_{z}n} \text{\Large $\Pi$} \left(\text{am}(nK(k) + \tau) ,-\tfrac{I_{z}(I_{x} - I_{y})}{I_{x}(I_{y} - I_{z})},k\right) + \Lambda
        \label{eq: closed_1}
        \end{equation}

$\triangleright \quad$ \textbf{Second case:} $2TI_{x}> G^{2} > 2TI_{y}>2TI_{z}$

    \begin{equation}
        \psi(K(k)) =  \frac{G}{I_{z}}K(k) - \frac{G(I_{x} - I_{z})}{I_{x}I_{z}n} \text{\Large $\Pi$} \left(\text{am}(nK(k) + \tau) ,-\tfrac{I_{z}(I_{x} - I_{y})}{I_{x}(I_{y} - I_{z})}k^{2},k\right) + \Lambda.
    \label{eq: closed_2}
    \end{equation}

Therefore, it is possible to determine a set of values so that the herpolhode is a closed curve, which occurs when the following relationship is established

\begin{equation}
    \psi(K(k)) = 2\pi\lambda, \qquad \lambda \in \mathbb N.
    \label{controle}
\end{equation}

Note that due to the accumulation of discrepancies as the movement occurs, it is possible that the initial conditions of the precession, nutation, and intrinsic rotation eventually synchronize. On the other hand, to obtain control over when and how the synchronization occurs, the condition (\ref{controle}) becomes an accurate tool for determining such a phenomenon.

\subsection{Numerical Solution}

With the intention of implement the numerical solution of a system of differential equations the function \textit{ode} already implemented in \textit{Scilab 6.1.0} is going to be applied. This function performs the numerical integration of the system using a user-defined method. For efficient computational calculations, the Adams-Bashford multi-step integration method will be applied.

To execute the numerical integration, it is necessary to obtain the differential equation that compounds the physical situation in the form of a state vector. From the differential system (\ref{Euler_cases}) it is obtained relationships for $\omega_{x}$, $\omega_{y}$, and $\omega_{z}$. To obtain the differential equation of the Euler's angles equation (\ref{eq: sistema_analitico}) will be considered

\begin{equation}
\begin{bmatrix}
    \sin\theta\sin\varphi & \cos \varphi & 0\\
    \sin\theta\cos\varphi & -\sin \varphi & 0\\
    \cos\theta &0&1
\end{bmatrix}
\begin{bmatrix}
    \dot{\psi}\\
    \dot{\theta}\\
    \dot{\varphi}
\end{bmatrix}
=
\begin{bmatrix}
    \omega_{x}\\
    \omega_{y}\\
    \omega_{z}
\end{bmatrix}
\label{eq: sistema_analitico_inicial}
\end{equation}

Therefore, solving the system (\ref{eq: sistema_analitico_inicial}) it is possible to obtain expressions for $\dot{\psi}$, $\dot{\theta}$ and $\dot{\varphi}$ as function of the Euler's angles and the angular velocity components.

Thus, to obtain the numerical solution for the angular velocity components and Euler's angles, the following differential system is considered

\begin{equation*}    \label{eq: não_linear}
    \begin{split}
        \dot{\psi} &= (\omega_{x}\sin\varphi + \omega_{y}\cos\varphi)\csc\theta\\
        \dot{\theta} &= \omega_{x}\cos\varphi - \omega_{y}\sin\varphi\\
        \dot{\varphi} &= \omega_{z} - \cot\theta(\omega_{x}\sin\varphi + \omega_{y}\cos\varphi)\\
        \dot{\omega}_{x} &= \frac{I_{y} - I_{z}}{I_{x}}\omega_{y}\omega_{z} \\
        \dot{\omega}_{y} &= \frac{I_{z} - I_{x}}{I_{y}}\omega_{x}\omega_{z} \\
        \dot{\omega}_{z} &= \frac{I_{x} - I_{y}}{I_{z}}\omega_{x}\omega_{y} 
        \end{split},
        \qquad
        \begin{split}
         \psi(0) &= 0\\
         \theta(0) &= \arccos \left(\frac{I_{z}\omega_{z0}}{G}\right)\\
         \varphi(0) &= \text{atan2} (I_{x}\omega_{x0},I_{y}\omega_{y0})\\
         \omega_{x}(0) &= \omega_{x0}\\
         \omega_{y}(0) &= \omega_{y0}\\
         \omega_{z}(0) &= \omega_{z0}
        \end{split}
\end{equation*}

\section{Results}

\subsection{Euler Poinsot Solver}

\begin{figure}[ht]
    \centering
    \includegraphics[width = \linewidth]{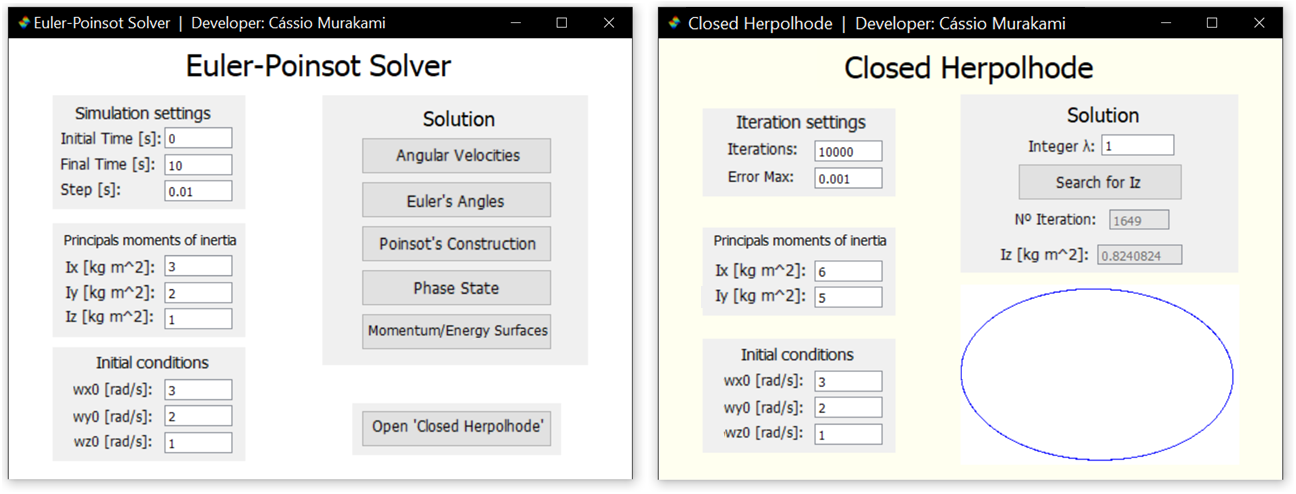}
    \caption{Euler-Poinsot Solver and Closed Herpolhode interfaces}
    \label{fig: GUI}
\end{figure}

In order to facilitate access to the results, a code was elaborated in the open-source software \textit{Scilab 6.1.0} that provides the solutions obtained in this paper for the Euler-Poinsot problem. Besides, a Graphical User Interface was implemented to create an intuitive user environment. The code is available at \textit{GitLab} \cite{link_Euler_Poinsot} and the implemented interface can be visualized in Figure \ref{fig: GUI}. 

The user inserts the simulation conditions, principal moments of inertia, and the components of the initial angular velocity. Thus, using the developed algorithms, the user obtains the solutions for the given imposed situation.

\begin{itemize}
    \item \textbf{Angular Velocities: } Display the analytical and numerical solutions for $\omega_{x}(t)$, $\omega_{y}(t)$, and $\omega_{z}(t)$.
    \item \textbf{Euler's Angles: } Display the analytical and numerical solutions for $\psi(t)$, $\theta(t)$, and $\varphi(t)$.
    \item \textbf{Poinsot's Construction: } Presents the animation of Poinsot's Geometric Solution.
    \item \textbf{Phase State: } Presents the Phase State of the situation.
    \item \textbf{Momentum/Energy Surfaces:} Presents the Momentum and Energy invariant ellipsoids.
\end{itemize}

By selecting \textit{Open 'Closed Herpolhode'} a new interface will be accessed in which, based on the theory developed in the dynamic model for the conditions in
that the herpolhode is a closed curve (\ref{eq: closed_1}) and (\ref{eq: closed_2}) returns the value of $I_{z}$ in which the closed herpolhode phenomenon occurs. The user inputs the parameters $I_{x}$, $I_{y}$, $\omega_{x0}$, $\omega_{y0}$, $\omega_{z0}$, and $\lambda$, which defines the multiple of $2\pi$ that
there is a synchronization between precession with intrinsic rotation and nutation.

\subsection{Simulation: $2TI_{x}>2TI_{y}>G^{2}>2TI_{z}$}

\begin{figure}[ht]
    \centering
    \includegraphics[width = \linewidth]{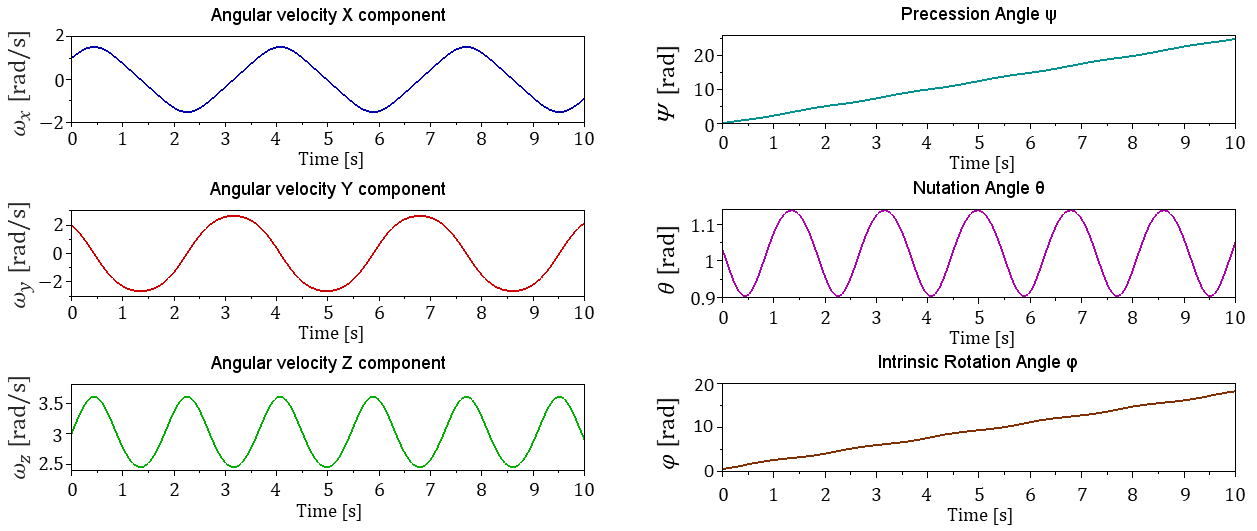}
    \caption{Angular velocities and Euler's angles results for the scenario $2TI_ {x}>2TI_{y}>G^{2}>2TI_{z}$}
    \label{fig: Solution_1}
\end{figure}

The following parameters were considered to study this case: $I_{x} = 3 \;\text{kg}\cdot\text{m}^2, I_{y} = 2 \;\text{kg}\cdot\text{m}^2,I_{z} = 1 \; \text{kg}\cdot\text{m}^2,\; \omega_{x}(0) = 1\;\text{rad/s},\;\omega_{y}(0) = 2\;\text{rad/s},\;\omega_{z}(0) = 3\;\text{rad/s}$. . The simulation contemplates the first 10 seconds of the motion, and the considered time step was 0.01 second. Figure \ref{fig: Solution_1} presents the result of the analytical solution of the scenario.

\begin{figure}[ht]
    \centering
    \includegraphics[width = 80 mm]{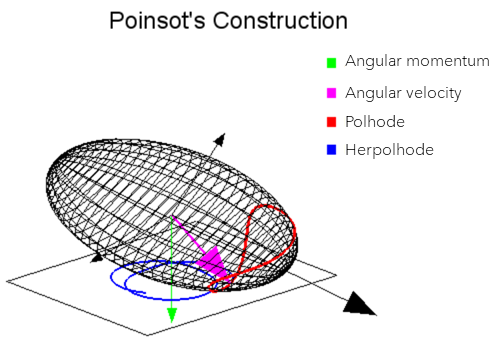}
    \caption{Poinsot's geometric solution animation for the scenario $2TI_ {x}>2TI_{y}>G^{2}>2TI_{z}$}
    \label{Poinsot_solution_1}
\end{figure}

The scenario presented a generic case of motion of the non-axisymmetrical rigid body around a fixed point, expressed by the elliptic functions described in (\ref{cases1}),  (\ref{euler_sol_1}), and (\ref{precession_sol_1}). The mean squared error ($\sigma$) between the analytical and numerical solution is negligible compared to the dimension of the other parameters due to numerical errors: $\sigma(\omega_{x}) = 7\times10^{-7}$, $\sigma(\omega_{y}) = 1\times10^{-6}$, $\sigma(\omega_{z}) = 6\times10^{-7}$, $\sigma(\psi) = 2\times10^{-2}$, $\sigma(\theta) = 5\times10^{-7}$, $\sigma(\varphi) = 2\times10^{-6}$. Therefore, is possible to conclude that results obtained in the analytical and numerical solutions for the angular velocities and Euler's angles coincided, which increases the reliability of the solutions \cite{thacker2004concepts}. Figure \ref{Poinsot_solution_1} represents an instant of time in the animation of the geometric solution using the results obtained by the analytical functions. 

The motion of Poinsot's ellipsoid was as expected since the phase space obtained agrees with the generated polhode. Also, the polhode includes the axis of least inertia, an expected condition for the initial conditions chosen.

\subsection{Simulation:  $2TI_ {x}>G^{2}> 2TI_{y}>2TI_{z}$}

\begin{figure}[h]
    \centering
    \includegraphics[width = \linewidth]{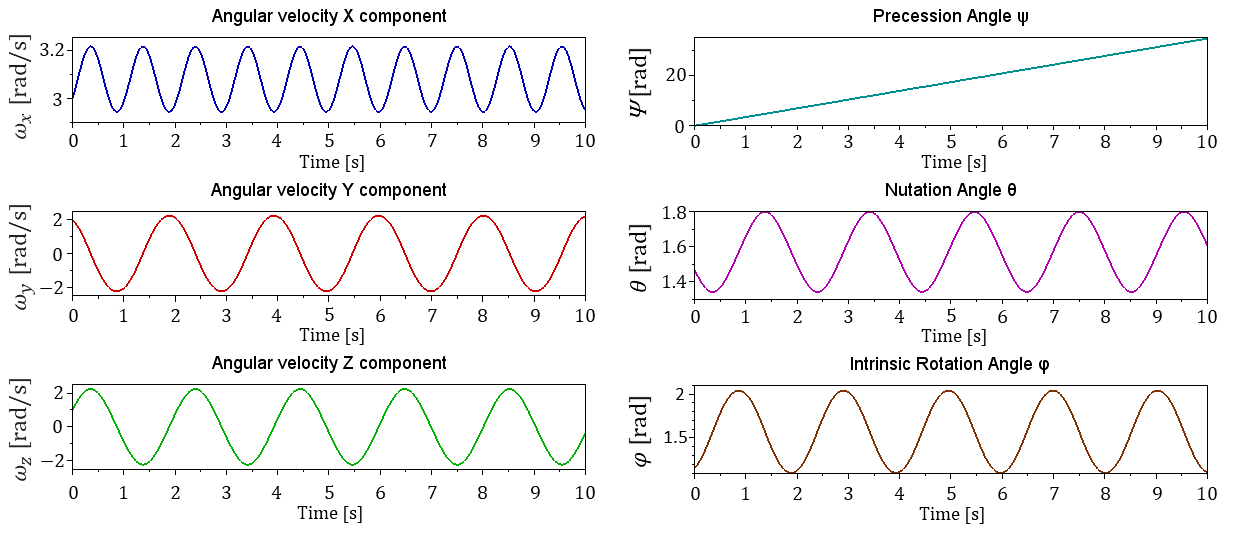}
    \caption{Angular velocities and Euler's angles results for the scenario $2TI_ {x}>G^{2}> 2TI_{y}>2TI_{z}$}
    \label{fig: Solution_2}
\end{figure}

The following parameters were considered to study this case: $I_{x} = 3 \;\text{kg}\cdot\text{m}^2,I_{y} = 2 \; \text{kg}\cdot\text{m}^2,I_{z} = 1 \; \text{kg}\cdot\text{m}^2,\; \omega_{x}(0) = 3\;\text{rad/s},\;\omega_{y}(0) = 2\;\text{rad/s},\;\omega_{z}(0) = 1\;\text{rad/s}$. The simulation contemplates the first 10 seconds of the motion, and the considered time step was 0.01 second. Figure \ref{fig: Solution_2} presents the result of the analytical solution of the scenario.

\begin{figure}[ht]
    \centering
    \includegraphics[width = 80 mm]{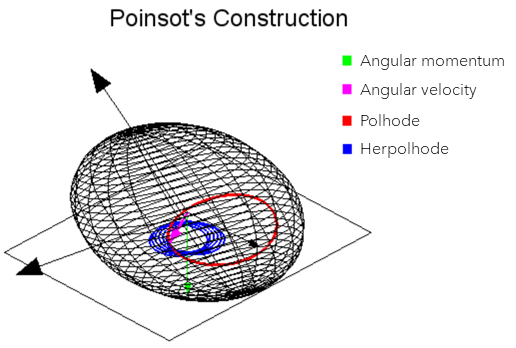}
    \caption{Poinsot's geometric solution animation for the scenario $2TI_ {x}>G^{2}> 2TI_{y}>2TI_{z}$}
    \label{Poinsot_solution_2}
\end{figure}

The scenario presented a generic case of motion of the non-axisymmetrical rigid body around a fixed point, expressed by the functions described in (\ref{cases2}),  (\ref{euler_sol_1}), and (\ref{precession_sol_2}). The mean squared error ($\sigma$) between the analytical and numerical solution is negligible compared to the dimension of the other parameters due to numerical errors: $\sigma(\omega_{x}) = 2\times10^{-7}$, $\sigma(\omega_{y}) = 1\times10^{-6}$, $\sigma(\omega_{z}) = 1\times10^{-6}$, $\sigma(\psi) = 4\times10^{-2}$, $\sigma(\theta) = 7\times10^{-7}$, $\sigma(\varphi) = 2\times10^{-6}$. Therefore, is possible to conclude that the results obtained in the analytical and numerical solutions for the angular velocities and Euler's angles also coincided, which increases the reliability of the solutions. Figure \ref{Poinsot_solution_2} represents an instant of time in the animation of the geometric solution using the results obtained by the analytical functions.

The motion of Poinsot's ellipsoid was as expected since the phase space obtained agrees with the generated polhode. Besides, the polhode includes the axis of greater inertia, an expected condition for the initial conditions chosen.

\newpage

\subsection{Closed herpolhode patterns}

Using the algorithm developed to obtain closed herpolhodes, two scenarios were analyzed that contemplate different formats of herpolhode. For the case where $2TI_{x} > 2TI_{y}>G^{2}> 2TI_{z}$ the following numerical values were chosen for the parameters: $I_{x} = 6 \; \text{kg}\cdot\text{m}^2,\;I_{y}=5 \; \text{kg}\cdot\text{m}^2,\; \omega_{x}(0) = 1\;\text{rad/s},\;\omega_{y}(0) = 2\;\text{rad/s},\;\omega_{z}(0) = 3\;\text{rad/s}$. For the case where $2TI_{x}>G^{2}>2TI_{y} >2TI_{z}$ the following parameters were adopted: $I_{x} = 6 \; \text{kg}\cdot\text{m}^2,\;I_{y}=5 \; \text{kg}\cdot\text{m}^2,\; \omega_{x}(0) = 3\;\text{rad/s},\;\omega_{y}(0) = 2\;\text{rad/s},\;\omega_{z}(0) = 1\;\text{rad/s}$. 

\begin{figure}[ht]
    \centering
    \includegraphics[width = \linewidth]{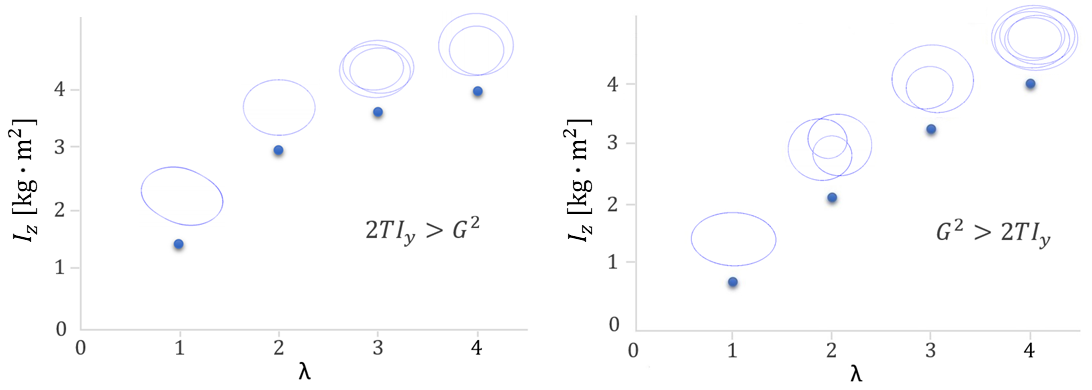}
    \caption{Closed herpolhode patterns observed for a given parameter $\lambda$ in a described scenario}
    \label{fig: Closed_Herpolhode_pattern}
\end{figure}

Subsequently, the images of the herpolhode obtain for certain $I_{z}$ associated with the parameter $\lambda$ were arranged in the graph illustrated in Figure \ref{fig: Closed_Herpolhode_pattern}.

\section{Conclusions}

Throughout the project, solutions were developed for the classic Euler-Poinsot problem. The first solution consists of the elaboration of the dynamic model based on the analytical solution of the system of differential equations. The second solution was based on the computational model, in which the differential equations were solved using numerical methods. Both results converged to the same outcome, as can be seen from the mean squared error. Such an experiment is a technique of validation called comparison with other models, which is used to increase confidence in the model.

Furthermore, the animation of Poinsot's geometric solution was elaborated, in a way that when comparing the result obtained by the analytical solution for the time evolution of the angular velocity with the polhode generated from the geometric solution it is possible to observe that both coincide. Therefore, once a reliable result was obtained for Poinsot's geometric solution, it is possible to simulate with precision the motion of a rigid body fixed at a point without the performance of external torques. It is worth mentioning that the study in the conditions so that herpolhode is a closed curve resulted in successful outcomes, due to the correct implementation of all previous results, which is a strong indication of the validity of the elaborated models.

To sum up, the methods of verification and validation of models were used, such as animation, comparison to other models, and operational graphics. The success of the model developed when subjected to tests reveals strong reliability in the representation of the studied phenomenon. Therefore, is possible to conclude that the mathematical development described in the paper and the implemented computational model are trustworthy.

\section{Acknowledgement}

This paper and the research behind it would not have been possible without the exceptional support of my supervisor, Professor Flavius Portella Ribas Martins. He provided insights and expertise that greatly assisted the research, also he offered an opportunity to participate in  undergraduate research which was the base for the elaboration of this paper.  I would also like to extend my gratitude to Ivaïlo Mladenov for the hospitality and his interest in this work and to the revisors of the article at the Journal of Geometry and Symmetry in Physics.

\bibliographystyle{unsrt}  
\bibliography{references} 

\begin{thebibliography}{10}

\bibitem{hazelrigg2003thoughts}
George~A Hazelrigg.
\newblock Thoughts on model validation for engineering design.
\newblock In {\em International Design Engineering Technical Conferences and
  Computers and Information in Engineering Conference}, volume 37017, pages
  373--380, 2003.

\bibitem{sargent2010verification}
Robert~G Sargent.
\newblock Verification and validation of simulation models.
\newblock In {\em Proceedings of the 2010 winter simulation conference}, pages
  166--183. IEEE, 2010.

\bibitem{andaloro1991modelling}
G~Andaloro, V~Donzelli, and RM~Sperandeo-Mineo.
\newblock Modelling in physics teaching: the role of computer simulation.
\newblock {\em International Journal of Science Education}, 13(3):243--254,
  1991.

\bibitem{garnier1954cinematique}
R.~Garnier and Facult{\'e} des sciences~de Paris.
\newblock {\em Cin{\'e}matique du point et du solide composition des
  mouvements}.
\newblock Cours de cin{\'e}matique. Gauthier-Villars, 1954.

\bibitem{meirovitch2010methods}
Leonard Meirovitch.
\newblock {\em Methods of analytical dynamics}.
\newblock Courier Corporation, 2010.

\bibitem{Goldstein}
Herbert Goldstein, Charles Poole, and John Safko.
\newblock {\em Classical Mechanics (3rd Edition)}.
\newblock Pearson, 06 2001.

\bibitem{zabunov2013effect}
Svetoslav~Svetoslavov Zabunov.
\newblock Effect of poinsot construction in online stereo 3d rigid body
  simulation on the performance of students in mathematics and physics.
\newblock {\em International Journal of Physics \& Chemistry Education},
  5(2):111--119, 2013.

\bibitem{Mladenov}
Clementina Mladenova and Ivailo Mladenov.
\newblock Spacecraft dynamics under the influence of gravity tourques.
\newblock {\em J. Theor. Appl. Mechanics}, 38:3--22, 09 2008.

\bibitem{greenhill1892applications}
George Greenhill.
\newblock {\em The applications of elliptic functions}.
\newblock Macmillan and Company, 1892.

\bibitem{landau1976mechanics}
LD~Landau and EM~Lifshitz.
\newblock Mechanics third edition: Volume 1 of course of theoretical physics.
\newblock {\em Elsevier Science}, pages 116--122, 1976.

\bibitem{byrd2013handbook}
Paul~F Byrd and Morris~D Friedman.
\newblock {\em Handbook of elliptic integrals for engineers and physicists},
  volume~67.
\newblock Springer, 2013.

\bibitem{analytic_formula}
Nicholas Mecholsky.
\newblock Analytic formula for the geometric phase of an asymmetric top.
\newblock {\em American Journal of Physics}, 87:245--254, 04 2019.

\bibitem{link_Euler_Poinsot}
Cássio Murakami.
\newblock {Euler Poinsot Solver},
  {\url{https://gitlab.com/cassiomura/euler-poinsot-solver}}, {Accessed:
  2021-03-09}.

\bibitem{thacker2004concepts}
Ben~H Thacker, Scott~W Doebling, Francois~M Hemez, Mark~C Anderson, Jason~E
  Pepin, and Edward~A Rodriguez.
\newblock Concepts of model verification and validation.
\newblock Technical report, Los Alamos National Lab., 2004.

\end{thebibliography}

\end{document}